\documentclass{aa}  

\usepackage{graphicx}
\usepackage{txfonts}
\usepackage{subcaption}       
\usepackage{lscape}            
\usepackage{placeins} 

\usepackage{orcidlink}
\usepackage{xcolor}
\usepackage{soul}

\usepackage{tikz}
\usetikzlibrary{arrows.meta, positioning, fit}
\usepackage{cancel}

\newcommand{\nmt}{{\tt NaMaster}}
\newcommand{\healpix}[0]{\textsc{HEALPix}}

\usepackage{hyperref}
\hypersetup{
    colorlinks=true,
    citecolor=blue,  
    linkcolor=blue,  
    urlcolor=blue    
}

\begin{document}

   \title{Probing dark energy evolution with Quaia quasars through the integrated Sachs–Wolfe effect}

   \author{Mina Ghodsi Yengejeh,
          \inst{1,2,3}\orcidlink{0000-0001-5481-9810}
            \and
        Jose R. Bermejo-Climent, \inst{1,2}\orcidlink{0000-0002-5825-579X}
            \and
        András Kovács, \inst{1,2}\orcidlink{0000-0002-5825-579X}
            \and
        Istv\'an Szapudi,
        \inst{4}\orcidlink{0000-0003-2274-0301}
           \and
        Istv\'an Csabai
        \inst{3}\orcidlink{0000-0001-9232-9898}
          }

    \institute{MTA--CSFK \emph{Lend\"ulet} ``Momentum'' Large-Scale Structure (LSS) Research Group, Konkoly Thege Mikl\'os \'ut 15-17, H-1121 Budapest, Hungary
    \and
    Konkoly Observatory, HUN-REN Research Centre for Astronomy and Earth Sciences, Konkoly Thege Mikl\'os \'ut 15-17, H-1121 Budapest, Hungary
    \and
    Institute of Physics and Astronomy, ELTE E\"otv\"os Lor\'and University, P\'azm\'any P\'eter s\'et\'any 1/A, H-1117 Budapest, Hungary
    \and
    Institute for Astronomy, University of Hawaii, 2680 Woodlawn Drive, Honolulu, HI 96822, USA
    }
    
   \date{Received Month Day, 2026; accepted Month Day, Year}

  \abstract
   {The Integrated Sachs--Wolfe (ISW) effect probes the late-time evolution of gravitational potentials and provides a complementary test of the nature of dark energy.}
   {We investigate whether the redshift evolution of the ISW effect can provide new constraints on the time evolution of dark energy. We consider theoretical predictions for the standard $\Lambda$CDM cosmology and alternative $w_0w_a$CDM models favoured by recent DESI BAO and DES Y6 constraints.}
   {We perform a tomographic cross-correlation analysis of the {\it Quaia} quasar catalogue and \textit{Planck} CMB temperature maps to measure the ISW signal over a broad redshift range, spanning about $(10h^{-1}~\mathrm{Gpc})^3$ comoving volume. We assess the robustness of the inferred ISW amplitude against variations in sky coverage, multipole range, and tomographic binning.}
   {We detect the ISW effect at a significance of $2.8 \, \sigma$, corresponding to an amplitude of $A_{\rm ISW}\simeq1.69\pm0.61$ relative to the \textit{Planck} $\Lambda$CDM prediction. The inferred signal remains stable against various analysis choices, including various CMB maps (SMICA, NILC, SEVEM), different choices of $\ell_{\rm max}$ and the number of multipole bins, and an alternative quasar bias model.}
   {The measured ISW amplitude is moderately stronger than predicted by the fiducial $\Lambda$CDM cosmology, and the alternative $w_0w_a$CDM models do not account for this discrepancy. Future tomographic ISW measurements with improved quasar catalogues from {\it Gaia} and forthcoming wide-area galaxy surveys such as {\it Euclid} and DESI will help clarify the origin of this difference.}


   \maketitle

\nolinenumbers

\section{Introduction}
\label{sec:intro}

The evolution of the large-scale structure of the Universe provides a unique opportunity to probe the physical properties of the dark sector, which accounts for approximately $95\%$ of the total cosmic energy density. In the standard $\Lambda$CDM model, cold dark matter (CDM) governs the growth of structure, while a cosmological constant, $\Lambda$, drives the observed late-time accelerated expansion of the Universe \citep{RiessEtal98}. 

Although $\Lambda$CDM provides an excellent description of a wide range of cosmological observations, the physical nature of dark energy remains unknown. This has motivated extensions in which the dark energy equation-of-state (EoS) parameter differs from, or evolves away from, the cosmological constant value of $w=-1$. The simplest such extension, $w$CDM, assumes a constant but free EoS parameter, whereas the widely adopted Chevallier--Polarski--Linder (CPL) parameterisation describes a time-dependent EoS, $w_0w_a$CDM,
where $a$ is the cosmic scale factor, and $w_0$ and $w_a$ characterize the present-day value and time evolution of the dark energy EoS, respectively \citep{Chevallier:2000qy,Linder:2002et}.

\begin{figure*}
\centering
\includegraphics[width=160mm]{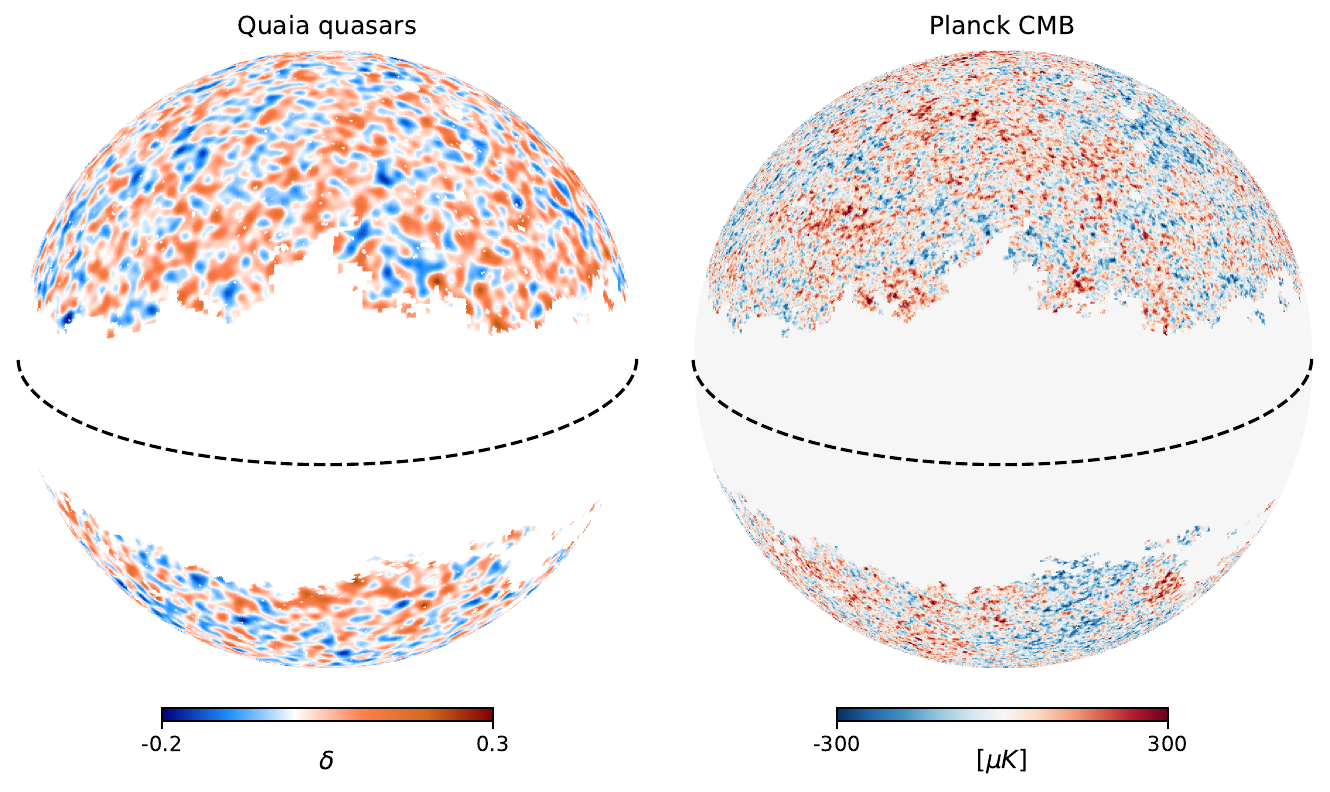}
\caption{Orthographic (half-sky) projections of the two maps entering our cross-correlations, left) {\it Quaia} quasar density map for the total sample smoothed with a Gaussian with FWHM $= 2 ^{\circ}$ to enhance the visibility of large-scale structure, and right) \textit{Planck} SMICA map in $\mu K$. Both maps are shown in Galactic coordinates and multiplied by the same combined binary mask. }
\label{fig:maps_view}
\end{figure*}

Distinguishing between different dark energy scenarios has become increasingly important as recent cosmological observations have begun to show intriguing hints in favour of a dynamical dark energy component. In particular, the latest baryon acoustic oscillation (BAO) measurements from the Dark Energy Spectroscopic Instrument (DESI), combined with weak-lensing and galaxy clustering constraints from the Dark Energy Survey (DES), as well as Type Ia supernova and cosmic microwave background (CMB) observations, exhibit a growing preference for departures from the cosmological constant paradigm \citep{DESI:2025zgx, DESI:2024mwx, DES:2025sig, DES:2026jmi}.

While different dark energy models can produce similar expansion histories, they generally predict different growth histories for large-scale structure and the associated evolution of gravitational potentials. As a consequence, they also predict distinct Integrated Sachs--Wolfe (ISW) signals \citep{Sachs1967}. The ISW effect arises as a secondary CMB anisotropy: as CMB photons travel from the surface of last scattering to the observer, they gain or lose net energy while traversing gravitational potentials that evolve with cosmic time, i.e. hot and cold spots are imprinted aligned with superclusters and voids, respectively \citep[see e.g.,][]{RudnickEtal2007,Granett2008,Cai2009,Hotchkiss2015, SzapudiEtAl2014}. The CMB therefore acts as a backlight that illuminates the intervening large-scale structure, making the ISW effect a unique and independent probe of dark energy.

Detecting the ISW effect is nevertheless challenging because its contribution is roughly an order of magnitude smaller than the primary CMB temperature anisotropies on the largest angular scales ($\ell\lesssim100$) and is fundamentally limited by cosmic variance. Consequently, ISW measurements rely on statistical cross-correlations between CMB temperature maps and large-scale structure tracers. The strength of the detected signal is commonly expressed through the amplitude parameter
\begin{equation}
A_{\rm ISW}:=
\dfrac{\mathrm{Measured\ ISW\ signal}}
{\mathrm{Predicted\ ISW\ signal}},
\label{Eq:A_ISW_text}
\end{equation}
where $A_{\rm ISW}=1$ corresponds to a baseline $\Lambda$CDM prediction.

Using this formalism, numerous ISW detections have been reported by cross-correlating CMB temperature maps with galaxy, quasar, and weak-lensing surveys \citep[see e.g.,][and references therein]{Boughn:2003yz, gianEtal08,ho, Xia:2009dr,PlanckISW2015,Stolzner2018,Hang2021,BahrKalus_2022,Krolewski:2021znk}. Their statistical significance (see Table \ref{tab:others_ISW_amplitudes}) depends primarily on the surveyed cosmic volume: wide sky coverage is required to sample the largest angular scales, while broad redshift coverage enables tomographic studies of the epoch of dark-energy domination. Even for an ideal full-sky tracer, the maximum achievable significance in $\Lambda$CDM is only $\sim7.5\,\sigma$, fundamentally limited by cosmic variance \citep{Afshordi_2004}; realistic surveys are further degraded by finite sky coverage, limited depth, and shot noise.

\begin{table}[ht]
\centering
\caption{Summary of recent $A_{\rm ISW}$ measurements, highlighting previous detections using combined analyses, radio surveys with wide N(z), and also QSO-based results in the high-$z$ range.}
\begin{tabular}{l|c|c|c}
First author & Year & Data & $A_{\rm ISW}$ \\ 
\hline \hline
Xia   & 2009 & QSO ($z < 1.5$) & 2.06 $\pm$ 0.75 \\
Planck Coll.   & 2016 & Combined ($z < 3.5$) & 1.00 $\pm$ 0.25 \\
Stölzner & 2018 & Combined ($z < 6.0$) & 1.51  $\pm$ 0.30 \\
 &   & QSO ($0.5 < z < 3.0$) & 2.41  $\pm$ 1.13 \\
 &   & NVSS ($z < 6.0$) & 1.70  $\pm$ 0.57 \\
Hang   & 2021 & DESI LS ($z \leq 0.8$) & 0.98 $\pm$ 0.35 \\
Bahr-Kalus   & 2022 & ASKAP ($z \lesssim 5.2$) & 0.94 $\pm$ 0.42 \\
Krolewski   & 2022 & unWISE ($z < 2.0$) & 0.96 $\pm$ 0.30 \\
\end{tabular}
\label{tab:others_ISW_amplitudes}
\end{table}

In this context, the {\it Quaia} catalogue of 1.3 million quasars \citep{Storey_Fisher_2024} provides a particularly attractive data set for ISW measurements. Although quasi-stellar objects (QSOs) are sparser tracers than galaxies (increasing the shot-noise contribution), {\it Quaia} combines nearly full-sky coverage with an exceptionally broad redshift range extending to $z\sim4$, i.e. spanning much of cosmic history since dark energy has been on the rise. 
In particular, our goal is to assess whether current ISW tomography has sufficient sensitivity to distinguish the $\Lambda$CDM prediction from the evolving dark-energy models recently favoured by DESI and DES.

The paper is organized as follows. In Sect.~\ref{sec:methods}, we describe the input data sets and our methodology. In Sect.~\ref{sec:results}, we present our main results, which are then further discussed in Sect.~\ref{sec:discussion}.

\section{Data and Methodology}
\label{sec:methods}

\subsection{CMB map}
For the CMB temperature map, we use the \textit{Planck} PR3 (2018) full-mission data release (see Fig.~\ref{fig:maps_view}, right). As our baseline foreground CMB map, we adopt the SMICA; we also use NILC and SEVEM as alternatives to check robustness. All maps are provided in \healpix \footnote{\url{http://healpix.sourceforge.net}} at $N_{\rm side} = 2048$ and are degraded to $N_{\rm side} = 256$ to match the resolution of our tracer maps \citep{gorski2005healpix, Zonca2019}. This has no impact on the ISW analysis, whose signal is mostly dominated in $\ell \lesssim 100$. Correspondingly, we use the \textit{Planck} "Component Separation Common mask in Intensity." To repixelise it consistently, we downgrade to the same $N_{\rm side} = 256$, and then make it a binary mask by keeping only pixels with coverage $> 0.99$ to prevent boundary leakage from the Galactic plane and point sources.

\subsection{Quaia quasar catalogue}
\label{subsec:quaia-catalog}
{\it Quaia} \citep{Storey_Fisher_2024} is a quasar catalogue that was constructed from the {\it Gaia} DR3 quasar candidates sample and unWISE \citep{Schlafly_2019} infrared data. It is a full-sky catalogue that covers the largest volume of any existing quasar sample, making it ideal for ISW measurements and other large-scale effects such as primordial non-Gaussianity (e.g., \citealt{Fabbian2025,Bermejo2026}). 

The complete sample, with $G < 20.5$ limiting magnitude, contains $1\,295\,502$ sources. There is a cleaner sample with $G < 20.0$ limiting magnitude and 755,850 quasars. The spectro-photometric redshifts were improved by training a k-nearest neighbors model on SDSS redshifts, achieving estimates on the $G < 20.0$ sample with only $\sim10\%$ catastrophic redshift errors, such that $\lvert \Delta z/(1+z)\rvert > 0.1$. In this paper, we used the $G < 20.5$ sample, for which $\sim70\%$ of the sources agree to $\lvert \Delta z/(1+z)\rvert < 0.1$ and $\sim$62\% to $\lvert \Delta z/(1+z)\rvert < 0.01$.

Together with the {\it Quaia} catalogue, the following materials are provided: a \healpix \, selection function map, which gives the probability that a source in a given pixel is included in the catalogue; a random catalogue, downsampled according to the selection function and containing about ten times the number of {\it Quaia} quasars; and observational systematics templates. All the products are publicly available\footnote{\url{https://zenodo.org/records/10403370}}. 

\begin{figure}
\centering
\includegraphics[width=85mm]{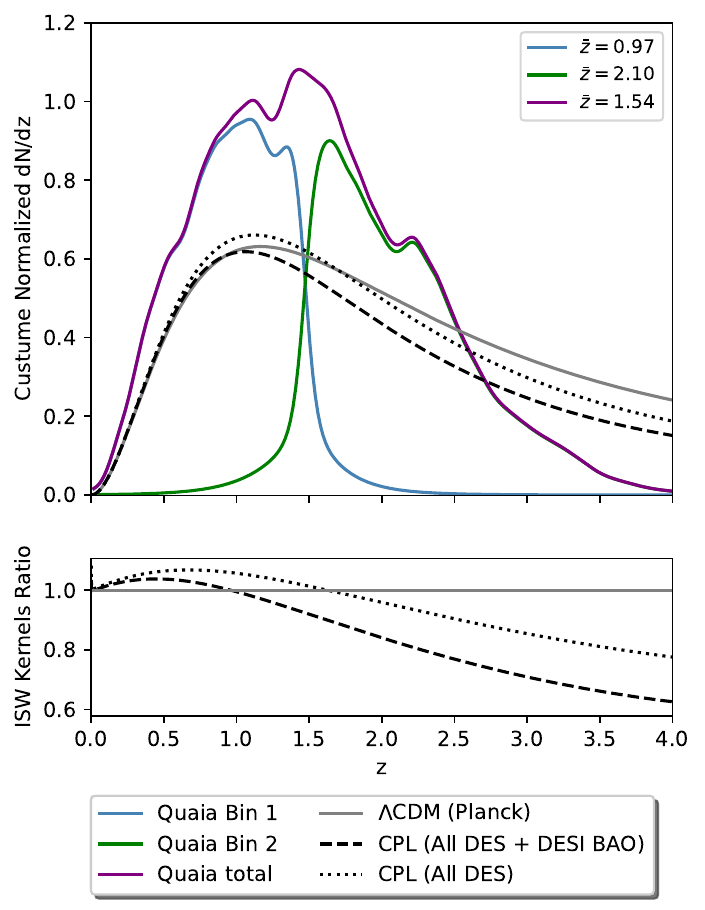}
\caption{Top panel: Normalised {\it Quaia} redshift distributions (dN/dz) for Bin 1 (blue), Bin 2 (green), and the total sample (purple), together with the ISW kernel per unit redshift ($W_{\rm ISW}(z) / H(z)$) computed with \texttt{pyCCL} for our three theoretical models: $\Lambda$CDM with the \textit{Planck} 2018 best-fit parameters (solid), CPL with the All DES + DESI BAO best-fit parameters (dashed), and CPL with the All DES best-fit parameters (dotted). Bottom panel: Ratio of the two CPL ISW kernels to the $\Lambda$CDM kernel over the range of $0 \leq z \leq 4.0$}.
\label{fig:redshift_dist}
\end{figure}

In this analysis, we will consider also the possibility of splitting the {\it Quaia} $G < 20.5$ sample in two redshift bins as in previous studies \citep{Alonso_2023,Fabbian2025,Bermejo2026}. The two redshift bins are equally populated, splitting the sample around $z \sim 1.5$ with average redshifts $\bar{z} = 0.97, \bar{z} = 2.10$. To account for the redshift uncertainties, we obtained the redshift distributions as the sum at every redshift of Gaussian photo-$z$ functions for each quasar given its observed redshift and redshift error. We represent in Fig.~\ref{fig:redshift_dist} the redshift distribution of the two $G< 20.5$ redshift bins together with the full {\it Quaia} distribution and the ISW kernels redshift dependence. 

To obtain the {\it Quaia} overdensity maps (see Fig.~\ref{fig:maps_view}, left), we first created quasar counts maps from the {\it Quaia} catalogue with $N_{\rm side} = 256$. The maps were created directly in Galactic coordinates, and we rotated the selection function from celestial to Galactic coordinates using the \texttt{healpy.rotator} package. Then, we defined the quasar overdensity map as the counts map corrected by the selection function, normalised by the mean quasar density and mean-subtracted. We imposed the selection function to be larger than 0.5 to exclude low completeness regions close to the Galactic plane, which could introduce systematics, and also masked out the Magellanic Clouds regions. Based on this threshold, we created a binary mask to select the pixels that enter the analysis.

We assumed a fiducial redshift evolution of the galaxy bias as measured in \citet{Piccirilli_2024} for the {\it Quaia} sample. This evolution is parameterised through the expression
\begin{equation}
\label{eq:bzquaia}
b_{\rm QSO} (z) = b_0 / D(z)  \,,
\end{equation}
with $b_0$ = 1.26 as fiducial value and $D(z)$ as the linear growth factor. As a robustness test, we also considered an alternative redshift evolution of the QSO bias (bias model $2$) based on the model by \citet{Laurent_2017}:
    \begin{equation}
    \label{eq:bz}
        b_{\rm QSO} (z) = 0.278 \left( (1+z)^2 - 6.565\right) + 2.393  \,.
    \end{equation}
For the magnification bias, in \cite{Alonso_2023} they measured a value $s = 0.404 \pm 0.004$  for the {\it Quaia} $G < 20.5$ limiting magnitude sample. This is compatible with the $s = 0.4$ value that would neglect the contribution from lensing magnification and other relativistic terms in the galaxy counts. Hence, in our theory estimation we neglected the contribution from these terms.

\subsection{Modelling tools}
We compute all theoretical angular power spectra with the Core Cosmology Library \citep[\texttt{pyCCL,}][]{Chisari_2019}, which evaluates the Limber approximation projection
\begin{equation}
    C^{X \, Y}_{\rm \ell} = \int W_{\rm X}(\chi) \, W_{\rm Y}(\chi) P_{\rm m} \big(k = \dfrac{\ell + 1/2}{\chi}, \, z(\chi) \big) \dfrac{d \chi}{\chi^2},
\end{equation}
where $\chi$ is the comoving distance, $W_{\rm X}(\chi)$ is the kernel of tracer $X$, and $P_{\rm m}(k, z)$ is the matter power spectrum calculated from \texttt{CAMB} \citep{2011ascl.soft02026L}, together with the halofit implementation \citep{Takahashi2012}. For each of our cosmological models and each {\it Quaia} redshift bin, we compute the ISW-quasar cross-spectrum $C^{Tg}_{\ell}$ over the multipole range of $2 \leq \ell \leq 767$. 

Our baseline model is a flat $\Lambda$CDM with the \textit{Planck} 2018 best-fit parameters \citep{Planck2018}. To study the sensitivity of the ISW cross-correlation to dynamical dark energy models, we additionally consider models with CPL parameterisation with an evolving EoS as
\begin{equation}
    w(a) = w_{\rm 0} + w_{\rm a} (1-a).
\end{equation}
Here we use two sets of parameters taken from the DES Year 6 analysis \citep[their Table S1]{DES:2026jmi}, "All DES + DESI BAO" ($w_0 = -0.84$, $w_a = -0.53$) and "All DES" ($w_0 = -0.84$, $w_a = -0.44$). Both sets of parameters favour a deviation from a cosmological constant that grows toward low redshift, where the ISW signal is sensitive.

Here we use two tracer kernels, the radial ISW kernel
\begin{equation}
    W_{\rm ISW} (\chi) \propto H(z) \, \chi^2 \, [1-f(z)],
\end{equation}
where $f(z)$ is the linear growth rate. 
The top panel of Fig.~\ref{fig:redshift_dist} shows this kernel, $W_{\rm ISW} (\chi) / H(z)$, for the $\Lambda$CDM (solid grey) and the two $w_0w_a$CDM models (black dashed and dotted), while the bottom panel shows the ratio of each $w_0w_a$CDM kernel to the $\Lambda$CDM one. All three curves peak at $z < 1.5$, showing the sensitivity of the ISW signal to the low-redshift Universe, where dark energy causes a gravitational decay over time. While the differences between the three curves become more pronounced at high redshift, this is where both the ISW kernel and the number of sources in our sample are weakest, limiting their impact.

We note that on non-linear scales, the gravitational potentials evolve even in the matter-dominated era, sourcing an additional Rees-Sciama (RS) contribution \citep{ReesSciama}. 
Although the RS signal contributes about $\sim10\%$ to the linear ISW term at $\ell \geq 200$  in the temperature auto-correlation \citep{Cai2010}, it is further suppressed in cross-correlation, especially using a sparse tracer catalogue like {\it Quaia}.

We also handle the number-count kernel of the quasar sample,
\begin{equation}
    w_{\rm g} (\chi) \propto b(z) \dfrac{dN}{dz},
\end{equation}
where $b(z)$ is a linear bias and $dN / dz$ is the normalised redshift distribution of the quasar sample (see Fig.~\ref{fig:redshift_dist}). \texttt{CCL} decomposes the observed galaxy density into a) density term, b) redshift-space distortion (RSD), and c) magnification bias. In this study, we do not consider the RSD contribution.

\subsection{Mock catalogues}
\label{subsec:mock-cat}
We generate a suite of $2000$ correlated Gaussian realisations of the CMB temperature field and of the {\it Quaia} quasar overdensity field, using both the full sample and two redshift bins configuration. For this, we used as input the theory angular power spectra ($C_\ell^{TT}, C_\ell^{gg}, C_\ell^{Tg}$) for the baseline 2018 best-fit cosmology. 

In order to account for the observed extra power in the {\it Quaia} quasar autocorrelation $C_\ell^{gg}$ due to observational systematics and provide a realistic estimation of the covariance matrix, we also added to the input $C_\ell^{gg}$ a power law contribution to better reproduce the data, following the procedure described in \citet{Bermejo2026} (see their Eq. 20 and Tab. 1).

For each realisation, we generated with the \texttt{healpy.synalm} function a single set of spherical-harmonic coefficients from the $N_{\rm fields}(N_{\rm fields}+1)/2$ auto- and cross-correlation input power spectra, so that the temperature and quasar density $a_{\rm \ell m}$ share the correct joint statistics and the ISW cross-correlation is preserved. The $a_{\rm \ell m}$ coefficients are calculated up to $\ell_{\rm max} = 3\, N_{\rm side} - 1 = 767$ and then synthesised into \healpix \, maps with $N_{\rm side} = 256$. To the temperature map we add an independent Gaussian contribution from the \textit{Planck} temperature noise power spectrum. 

For the {\it Quaia} density maps, to simulate the shot noise component due to having a finite number of quasars, we constructed number counts maps by Poisson-sampling the expected number density of quasars in each pixel, given by $N_{\rm expected} = \bar{N} (\delta +1)$, where $\bar{N}$ is the mean number of quasars per pixel and $\delta$ the simulated overdensity field. Finally, we convert the quasar counts maps back to overdensity maps. The resulting $2000$ temperature and quasar map pairs are used later to estimate the covariance of the ISW cross-correlation measurements. 

\subsection{Cross-correlation approach}

We compute all angular cross-power spectra with the public code \nmt \footnote{\url{https://github.com/LSSTDESC/NaMaster}}, which uses a pseudo-$C_{\rm \ell}$ formalism \citep{Alonso:2018jzx}. On a masked sky, the direct pseudo-spectrum $\tilde{C}_{\rm \ell}$ is a biased estimator of the true $C_{\rm \ell}$, because the mask couples different multipoles. \nmt \, corrects this by constructing the mode-coupling matrix $M_{\rm \ell \ell'}$ from the harmonic transform of the mask and analytically deconvolving it, so that we can get an unbiased estimate of the $C_{\rm \ell}$ as:
\begin{equation}
    \langle \Tilde{C}_\ell \rangle = \sum_{\ell'} M_{\ell \ell'} C_{\ell'} \,.
\end{equation}
In \nmt, the inversion of the $M_{\ell \ell'}$ matrix is done using the MASTER algorithm \citep{Hivon_2002}, which requires a discrete binning of the angular power spectrum.

\begin{figure*}
\centering
\includegraphics[width=174mm]
{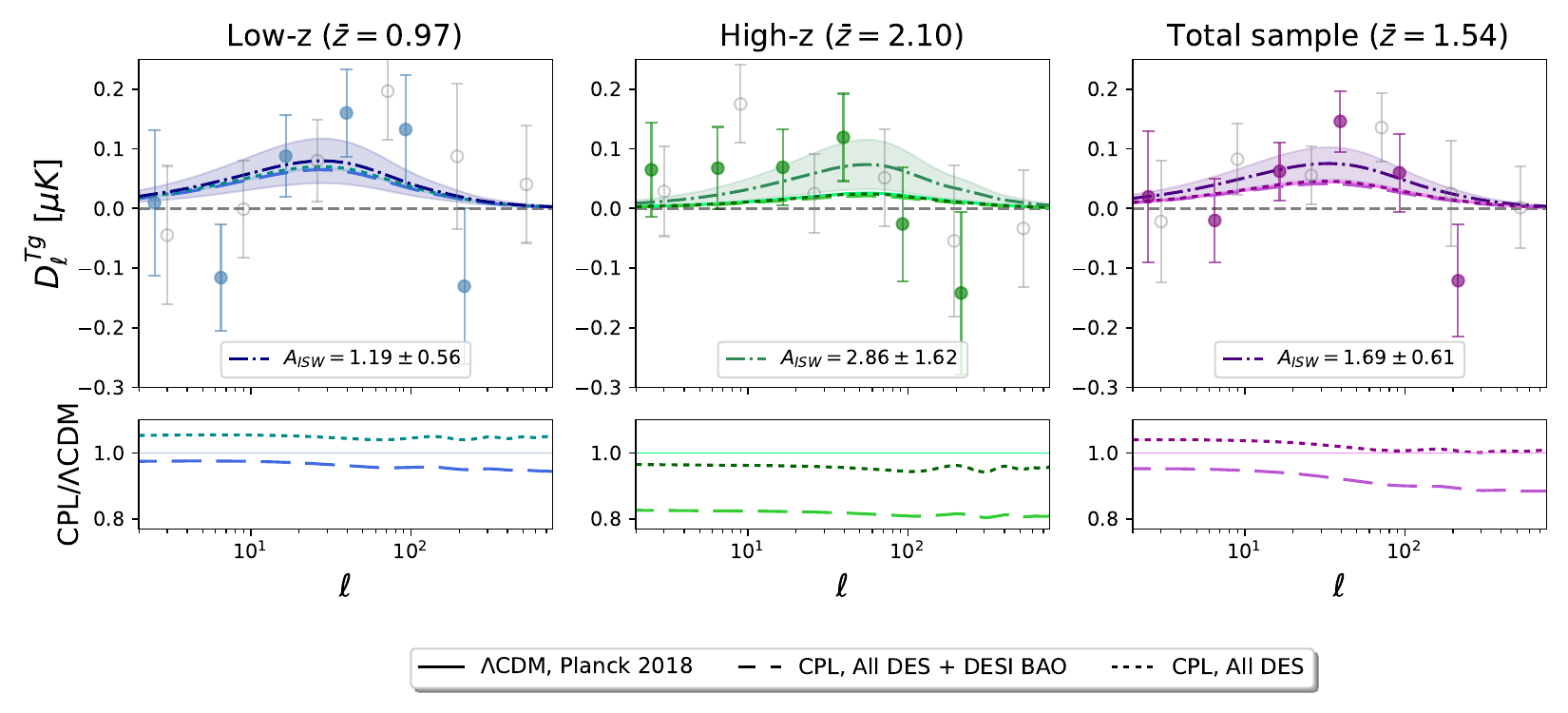}
\caption{Top row: Binned angular cross-power spectra $D_{\rm \ell}^{Tg} = \ell (\ell + 1) C^{Tg}_{\rm \ell} / (2 \pi)$ between the \textit{Planck} SMICA temperature map and the {\it Quai
a} quasar density maps, shown for the low-redshift Bin 1 (left), high-redshift Bin 2 (middle), and the total sample (right). Data points obtained with \nmt \, using six logarithmically spaced multipole bins up to $\ell_{\rm max} = 300$. Error bars are estimated from the standard deviation of the cross-correlation between the $2000$ random {\it Quaia} density maps and $2000$ random Gaussian CMB realisations generated from the \textit{Planck} 2018 temperature power spectrum. Solid coloured curves show the theoretical ISW cross-correlation prediction for the \textit{Planck} 2018 $\Lambda$CDM model, while the dashed curves show the predictions for the CPL best-fit cosmologies from All DES (short dashed) and All DES + DESI BAO (long dashed).
The coloured band shows the $\Lambda$CDM curve rescaled by the best-fit $A_{\rm ISW}$ with its $1\, \sigma$ uncertainty quoted in each panel legend.
Bottom row: ratio of the two $w_0w_a$CDM theory curves to the $\Lambda$CDM prediction of the corresponding column. 
Empty gray markers show the extended measurements using $6$ bins up to $\ell_{\rm max} = 767$, helping to interpret the nature of outlier data points in the fiducial analysis.}
\label{fig:3panel}
\end{figure*}

In our study, both CMB temperatures and quasar density maps entered as spin$-0$ quantities, and the mode-coupling matrices are precomputed once per mask configuration and reused for all cross-spectra sharing that mask pair. 
We adopt the conservative choice of taking the intersection of the two survey masks. This keeps only pixels where both surveys have reliable data and avoids introducing spurious cross-correlations from regions covered only by one map. We construct three joint masks: 
\begin{itemize}
    \item One for each {\it Quaia} redshift bin by multiplying the corresponding {\it Quaia} mask with the \textit{Planck} mask 
    ($\sim 24\,000 \, {\rm  deg}^2$ for low-$z$ bin 1 and the total sample, and $\sim 21\,000 \, {\rm deg}^2$ for high-$z$ bin 2). 
    \item To further test possible large-scale systematics associated with different hemispheres due to scanning patterns, we split each common mask mentioned above into Galactic Northern ($\theta < \pi/2$) and Southern ($\theta \geq \pi/2$) parts
    ($\sim 12\,000 \, {\rm deg}^2$ for low-z bin 1 and the total sample, and $\sim 10\,000 \, {\rm deg}^2$ for high-z bin 2). 
\end{itemize}

\subsection{Amplitude estimation and covariance}
\label{subsec:Amp_cov}
We estimate the ISW amplitude defined in Eq.~\ref{Eq:A_ISW_text} such that $A_{\rm ISW} = 1$ corresponds to the fiducial $\Lambda$CDM prediction, by fitting the binned theoretical cross-spectrum template $t_{\rm b} \equiv C_{\rm b}^{Tg, \, th}$ to the measured $d_{\rm b} \equiv C_{\rm b}^{Tg, \, data}$. Assuming a Gaussian likelihood, the amplitude is obtained by minimising
\begin{equation}
    \chi^2 (A_{\rm ISW}) = \sum (d_{\rm b} - A_{\rm ISW} t_{\rm b}) \, Cov^{-1}_{\rm bb'} \, (d_{\rm b'} - A_{\rm ISW} t_{\rm b'})
\end{equation}
where $Cov^{-1}_{\rm bb'}$ is the inverse of the covariance matrix. In this way, we will have
\begin{equation}
    A_{\rm ISW} = \dfrac{t^T Cov^{-1} d}{t^T Cov^{-1} t}, \hspace{1.5cm} 
    \sigma = \dfrac{1}{\sqrt{t^T Cov^{-1} t}},
\end{equation}
and we report the detection significance as $S/N \equiv A_{\rm ISW} / \sigma$. 

The covariance matrix is estimated from the sample variance of $2000$ realisations, generated by cross-correlating each random CMB map with each corresponding random {\it Quaia} map (see subsection \ref{subsec:mock-cat}). To correct the bias assumed by inverting a covariance matrix estimated from a finite number of realisations, we apply a correction derived by \cite{hartlap2007} as
\begin{equation}
    \dfrac{N_{\rm s} - N_{\rm b} -2}{N_{\rm s} - 1},
\end{equation}
where $N_{\rm s}$ is the number of realisations and $N_{\rm b}$ is the number of bins. In our fiducial setup, the data vector contains six bins, resulting in a correction factor of $0.996$.

Following \cite{cabre_etal}, we also tested an alternative, approximate error estimation method where we only randomized the CMB part and cross-correlated the 2000 CMB realisations with the {\it Quaia} data map. In that case, the true errors are underestimated by about $\sim 10 \%$ and a simple correction term must be added to the covariance matrix to account for this mismatch. We checked that our error estimation method and this approximate method both give consistent $A_{\rm ISW}$ results.

\section{Results}
\label{sec:results}

\subsection{Fiducial analysis}

\begin{table*}
\centering
\begin{tabular}{l|l|l|l}
Model & Bin 1 ($\bar{z} = 0.97$) & Bin 2 ($\bar{z} = 2.10$)& Total ($\bar{z} = 1.54$)\\
\hline
\hline 
$\Lambda$CDM ($A_{\rm ISW} = 1$) & $\chi^2 = 6.507$ & $\chi^2 = 5.925$ & $\chi^2 = 5.954$ \\
$\Lambda$CDM ($A_{\rm ISW}$ best-fit) & $\Delta \chi^2 = -0.115$ & $\Delta \chi^2 = -1.324$ & $\Delta \chi^2 = -1.283$ \\
No signal ($A_{\rm ISW} = 0$) & $\Delta \chi^2 = 4.414$ & $\Delta \chi^2 = 1.800$ & $\Delta \chi^2 = 6.417$ \\
CPL (DES) & $\Delta \chi^2 = -0.029$ & $\Delta \chi^2 = 0.054$ & $\Delta \chi^2 = -0.082$ \\
CPL (DES + DESI) & $\Delta \chi^2 = 0.072$ & $\Delta \chi^2 = 0.260$ & $\Delta \chi^2 = 0.282$ \\
\hline 
$w$CDM ($w=-0.65$) & $\Delta \chi^2 = 0.073$ & $\Delta \chi^2 = -1.443$ & $\Delta \chi^2 = -1.353$ \\
\end{tabular}
\caption{$\chi^2$ values and differences relative to the fiducial $\Lambda$CDM total case for fixed $A_{\rm ISW} = 1$ amplitude (first row, df=6), best-fit $A_{\rm ISW}$ (second row, df=5), no signal ($A_{\rm ISW} = 0$, df=6), two CPL model setups (df=6), and for a $w$CDM model (df=6). 
} 
\label{tab:chi-2}
\end{table*}

Our measured cross-power spectra between the \textit{Planck} SMICA temperature map and the {\it Quaia} quasar maps is shown in Fig.~\ref{fig:3panel}, split into the low-$z$ (left), high-$z$ (middle), and the total sample (right) panels. The binned coloured data points are computed with \nmt \, in six logarithmically spaced multipoles (i.e. 6 degrees-of-freedom) up to $\ell_{\rm max} = 300$, with the uncertainty calculated from the covariance of the cross-correlation between $2000$ random Gaussian CMB realisations drawn from the \textit{Planck} 2018 best-fit temperature spectrum and 2000 correlated mock {\it Quaia} maps. 

Fitting the $\Lambda$CDM curve to each $C_\ell^{Tg}$ measurement constrains best-fit ISW amplitudes of $A_{\rm ISW} = 1.19 \pm 0.56$ for the low-$z$ bin, $A_{\rm ISW} = 2.86 \pm 1.62$ for the high-$z$ bin, and $A_{\rm ISW} = 1.69 \pm 0.61$ for the total sample, corresponding to a detection of $2.13$, $1.77$, and $2.78$ $\sigma$ significance, respectively. The coloured bands in Fig.~\ref{fig:3panel} show the rescaled $\Lambda$CDM theory curve with the corresponding $A_{\rm ISW}$ and their uncertainties, showing best-fit amplitudes moderately above the fiducial model prediction. 

This slight ISW excess signal from the {\it Quaia} data fits a broader pattern in the literature \citep[see e.g.,][for pioneering work and follow-up analyses]{Granett2008,Hernandez2013,Ilic2013,cai2013,Hang2021}. For instance, \cite{Kovacs2019} used a combination of SDSS and DES data, and reported $A_{\rm ISW} \approx 5.2 \pm 1.6$ when performing CMB stacking analyses on the most extreme super-structures, large voids and superclusters. Analyzing the CMB {\it Cold Spot} area, \cite{Kovacs2022} confirmed the existence of the Eridanus supervoid in alignment, but based on the size and underdensity of the structure, a full explanation via $\Lambda$CDM ISW modeling is unlikely \citep[see e.g.][]{Nadathur2015,Naidoo2017}. 

There are also moderately significant findings about possible ISW sign-changes both from low-$z$ galaxy samples \citep[see e.g.,][]{Hansen:2025atx} and high-$z$ QSO catalogues \citep{Kovacs2021}, which further challenged the standard ISW models in recent years. The origin of these anomalous ISW results remains unclear, as more traditional $2$-point correlation measurements, typically recover $A_{\rm ISW} \approx 1$, including our analysis using QSOs \citep[see e.g.,][]{PlanckISW2015,Ferraro2015, NadathurCrittenden2016, Stolzner2018, Hang20212pt, BahrKalus_2022, Krolewski:2021znk}.

\subsection{Alternative dark energy models}

Alongside the fiducial $\Lambda$CDM model (solid lines), each top panel shows the theoretical $D^{Tg}_{\rm \ell} \equiv \ell(\ell+1)/ 2 \pi C_\ell^{Tg}$ for the $w_0w_a$CDM models obtained from All DES + DESI BAO (long dashed lines) and from All DES (short dashed lines). The bottom panels show the ratio of each alternative model to $\Lambda$CDM over the same multipole range. 
For the $w_0w_a$CDM theory curves, we first compute the dark matter auto-correlation $C^{mm}_{\ell}$, and then rescale the best-fit QSO bias values and set ${b_{0}, \,}_\mathrm{CPL} = {b_{0}, \,}_{\Lambda  \mathrm{CDM}} \times \sqrt{C^{mm}_{\Lambda \mathrm{CDM}} / C^{mm}_\mathrm{CPL}}$ and average over multipoles. This method changes the fiducial ${b_{0}, \,}_{\Lambda \mathrm{CDM}} = 1.26$ value to $1.32$ and $1.29$ for All DES + DESI BAO and All DES, respectively, as a consequence of slightly different clustering in these models.

Although the $A_{\rm ISW}$ values sit slightly above the $\Lambda$CDM expectation, the $w_0w_a$CDM models predict mostly a {\it lower} amplitude than $\Lambda$CDM (Fig.~\ref{fig:3panel}, bottom row), suppressed by at most $20\%$ across $2 \leq \ell \leq 300$. The deviation is larger for the All DES + DESI BAO fit in the high-redshift bin, but this is even much below the statistical uncertainty of the present {\it Quaia} measurements, so the current data cannot precisely distinguish between $\Lambda$CDM and the CPL parameterization. 
Nonetheless, we list the related statistical results in Table~\ref{tab:chi-2}, showing that none of the $w_0w_a$CDM models improves the fit to our measurements for the total sample ($\Delta \chi^2 = -0.082$ and $+0.282$). One may compare this with $\Delta \chi^2 = -1.283$ obtained from the rescaled $\Lambda$CDM model by the best-fit amplitude.

We also briefly explored other models in this context. For example,  \cite{GhodsiYengejeh:2025god} reported that quintessence models ($w>-1$) tend to predict an overall higher $A_{\rm ISW}$ amplitude compared to phantom models ($w<-1$), although this preference is partly degenerate with the assumed values of $\Omega_{\rm m}$ and $\sigma_8$, and depends on the redshift range considered. As an exploratory test, we computed the ISW prediction for a $w$CDM cosmology with $w = -0.65$ and ${b_{0}, \,}_{wCDM} = 1.32$, which raises the $D^{Tg}_{\rm \ell}$ well above $\Lambda$CDM and gives $\Delta \chi^2 = -1.353$, i.e. comparable to the rescaled $\Lambda$CDM model with $A_{\rm ISW} = 1.69 \pm 0.61$. While this ad-hoc choice for the $w$ parameter is plausibly ruled out by other existing observational constraints, it demonstrates the sensitivity of the ISW signal to the dark energy EoS.

\subsection{Robustness tests}
\begin{figure}
\centering
\includegraphics[width=85mm]{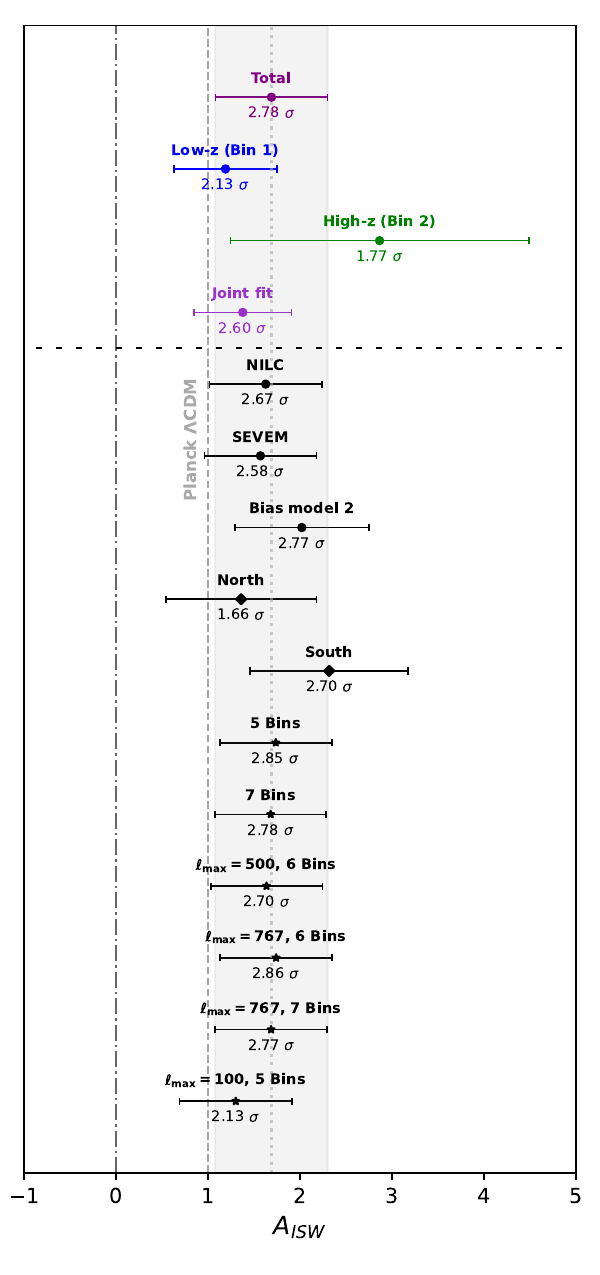}
\caption{The best-fit amplitude $A_{\rm ISW}$ for all analysis setups. Markers show central values with $1\, \sigma$ error bars, and numbers below show detection significance. The top three markers are our baseline SMICA $\times$ {\it Quaia} results, separated by a dashed line from robustness tests using alternative component-separation maps (NILC, SEVEM), a different bias model, North-South hemisphere splits, and different choices for binning and $\ell_{\rm max}$. The vertical dot-dashed line marks $A_{\rm ISW} = 0$, and the vertical dashed line marks the $\Lambda$CDM expectation $A_{\rm ISW} = 1$. The grey band shadows the $1\, \sigma$ range of the baseline total sample measurement.}
\label{fig:whisker}
\end{figure}

Fig.~\ref{fig:whisker} summarises the best-fit $A_{\rm ISW}$ amplitude calculated across all cross-correlation configurations considered in this analysis. The top four entries show our baseline measurements from the SMICA cross-correlation with the {\it Quaia} maps for the total sample (purple), the low-$z$ bin (blue), and the high-$z$ bin (green), corresponding to the three panels in Fig.~\ref{fig:3panel}. 

The fourth entry above the dashed line (light purple) shows a \emph{joint fit}, in which a single amplitude is constrained to a concatenated low-$z$ plus high-$z$ data vector. Here we used the full covariance between the 12 multipole bins, including correlations between the two redshift bins estimated from the same set of $2000$ random maps. The result is consistent with both individual redshift bins and with the total sample. 

To assess the robustness of the baseline result (shown by the horizontal dashed line), we performed a series of pipeline variations, and compared them against the $1\, \sigma$ range of the baseline results (grey band):

\begin{itemize}
    \item replacing the SMICA map with NILC and SEVEM versions slightly lowers the amplitudes to $A_{\rm ISW} = 1.63 \pm 0.61 \, (2.67 \, \sigma)$ and $A_{\rm ISW} = 1.57 \pm 0.61 \, (2.58 \, \sigma)$, respectively. In total, they show $0.2 \, \sigma$ shift with respect to the baseline. \smallskip
    \item adopting the alternative QSO bias evolution by \cite{Laurent_2017} raises the amplitude to $A_{\rm ISW} = 2.02 \pm 0.73$, which gives the significance essentially unchanged at $2.78 \, \sigma$. \smallskip
    \item splitting the {\it Quaia} maps into the North ($A_{\rm ISW} = 1.36 \pm 0.82$) and South ($A_{\rm ISW} = 2.32 \pm 0.86$) hemispheres show consistent values within their large statistical uncertainties, with each other and also with the fiducial result \citep[see][for details]{Gratton_2020}. \smallskip
    \item finally, varying the number of $\ell$-bins (5 or 7 instead of 6) and varying the maximum multipole ($\ell_{\rm max} = 100, 500$ and $767$) produces amplitudes in the range of $1.30$ to $1.74$, still within the $1\, \sigma$ band of the baseline result. 
\end{itemize}

Therefore, the ISW detection reported here is robust against the specific choices of CMB maps, galaxy bias model, North-South split, and the number of bins. The consistency between SMICA, NILC and SEVEM is particularly reassuring because these methods rely on substantially different foreground-removal strategies, implying that residual Galactic foregrounds or component-separation artefacts are unlikely to dominate the measured cross-correlation. 

While unresolved extragalactic emission associated with large-scale structure (e.g. the CIB, radio galaxies, or the tSZ effect) may also induce weak correlations with CMB temperature maps, these are expected to be subdominant on the large angular scales and at the current level of statistical precision.

We note that the measured $D^{Tg}_{\ell}$ does show some outlier data points; one at $\ell <10$ for the low-$z$ sample, and more importantly at $\ell \approx 200$ even for the total sample. The ISW signal is intrinsically cosmic-variance-limited, so this might explain low-$\ell$ fluctuations. In the high-$\ell$ regime, errors from both shot noise and the CMB data are higher, and at the same time theoretical models predict nearly zero ISW signal at small scales, especially in the high-$z$ bin, both suggesting a statistical fluctuation.

To further check possible systematic effects, we also extended the analysis to higher $\ell$, finding consistent $A_{\rm ISW}$ amplitudes. As shown in Fig.~\ref{fig:3panel}, we see different outlier data point patterns as we move from the fiducial analysis to an alternative binning scheme (in gray), again suggesting statistical fluctuations. 

In this context, since the nonlinear RS contribution is expected to be largest at these higher multipoles, the fact that our results remain unchanged confirms our measurements are not biased at current sensitivity. As a further test, we restricted the analysis to $\ell_{\rm max} = 100$ with $5$ bins, to be safely inside the linear regime, and again recovered a consistent amplitude, $A_{\rm ISW} = 1.30 \pm 0.61$.

\begin{table}[ht]
\centering
\caption{Summary of measured amplitudes $A_{\rm ISW}$ for $\ell_{\rm max} = 300$ and 6 bins, that is our fiducial measurement setup. See also Figs.~\ref{fig:3panel} and \ref{fig:whisker} for the related $A_{\rm ISW}$ analysis. 
}
\begin{tabular}{l|c|c|c}
Map & $\bar{z}$ & $A_{\rm ISW}$ & S/N\\
\hline \hline
Total sample  & 1.54 & 1.69 $\pm$ 0.61 & 2.78 $\sigma$	\\
Low-$z$ (Bin 1)  & 0.97	& 1.19 $\pm$ 0.56 & 2.13 $\sigma$	\\
High-$z$ (Bin 2)  &  2.10  & 2.86 $\pm$ 1.62 & 1.77 $\sigma$	\\
Joint fit  & 1.54  & 1.38 $\pm$ 0.53 & 2.60 $\sigma$	\\
\end{tabular}
\label{tab:ISW_amplitudes}
\end{table}

\section{Discussion and Conclusions}
\label{sec:discussion}

The {\it Quaia} quasar catalogue has been used for a range of studies of the cosmic large-scale structure. Its angular clustering combined with the cross-correlation with \textit{Planck} convergence has been used to constrain $\sigma_8 = 0.766 \pm 0.034$ and $\Omega_{\rm m} = 0.343^{+0.017}_{-0.019}$ \citep{Alonso_2023}. The same cross-correlation has measured the turnover of the matter power spectrum, detected at $2.3 - 3.1 \, \sigma$ and determining the matter-radiation equality scale at the $\sim 20\%$ level \citep{Alonso2025}. The catalogue has also been used to map the high-redshift cosmic web to construct a public void and cluster catalogue \citep{Arsenov:2025dse}. Related to our main interests, a recent analysis by \cite{Andrews-BORG} used {\it Quaia} to probe the large-scale gravitational potential. They applied a Bayesian field-level inference (BORG) to reconstruct the three-dimensional {\it Quaia} density field over a $\sim (10 \, h^{-1} \, \mathrm{Gpc})^3$ comoving volume, detecting its cross-correlation with the \textit{Planck} CMB lensing, a signal at $\sim 4 \, \sigma$ significance. 

Here we probed the decay of this potential due to dark energy, and measured the tomographic ISW signal. We cross-correlated the {\it Quaia} quasar catalogue with the \textit{Planck} CMB temperature map in two redshift bins and compared the result with the predictions of $\Lambda$CDM and $w_0w_a$CDM cosmologies favoured by recent DESI BAO and DES Y6 constraints. Our fiducial measurement gives $A_{\rm ISW} = 1.69 \pm 0.61$, a $2.78 \, \sigma$ detection consistent with the $\Lambda$CDM expectation of $A_{\rm ISW} = 1$ at $1.1\, \sigma$ (see Table \ref{tab:ISW_amplitudes} for more detail). This is in close agreement with the ISW amplitude  $A_{\rm ISW} = 1.70 \pm 0.57$ obtained by \cite{Stolzner2018} from the \texttt{NVSS} data set that also covers a wide redshift range. 

Considering subsets, our high-$z$ result $A_{\rm ISW} = 2.86 \pm 1.62$ at $\bar{z} = 2.10$ is consistent with $A_{\rm ISW} = 2.41 \pm 1.13$ reported for a quasar catalogue by \cite{Stolzner2018}, as well as with $A_{\rm ISW} = 2.06 \pm 0.75$ constrained by \cite{Xia:2009dr} using SDSS QSOs, covering a smaller area but with higher tracer density. Although this best-fit amplitude from {\it Quaia} is higher than expected in $\Lambda$CDM, the uncertainty is also substantially large because the ISW kernel is weaker and the tracer density lower at high redshifts, so the difference is not statistically significant.

Our results are robust against different variations we tested, swapping SMICA for NILC or SEVEM, using an alternative QSO bias model, splitting the sky into Northern and Southern hemispheres, and changing the maximum multipoles and number of bins. All of them shift the amplitude by less than $1\, \sigma$ of the fiducial model. In particular, restricting the fit to $\ell_{\rm max} = 100$ within the linear regime constrains an amplitude consistent with our fiducial result, suggesting that it is insensitive to non-linear RS contributions or possible small-scale systematics.

We also compared our measurements against two alternative $w_0w_a$CDM models based on DESI BAO + DES Y6; both predict mostly a lower $D^{Tg}_{\rm \ell}$ than $\Lambda$CDM across the multipole range ($\Delta \chi^2 = - \, 0.082$ and $+ \, 0.282$ for the total sample). However, we stress that the ISW signal alone cannot discriminate between these models, as the measurement errors are larger than differences in the theoretical predictions of CPL and $\Lambda$CDM models. At this point, the $A_{\rm ISW}$ constraints from the {\it Quaia} analysis thus serve as a consistency check. We note, however, that our results prefer an amplitude above the $\Lambda$CDM template, not below as predicted by the CPL models, motivating more detailed future modeling analyses. 

Above all, the tomographic ISW measurements presented here are currently limited not only by cosmic variance, but also by the low number density of quasars. Forthcoming surveys, such as \textit{Gaia} DR4, will increase the source density and extend the dataset to higher redshift, as well as \texttt{Euclid} and DESI, which will provide more detailed comsic maps over the $z \leq 1.5$ range where the ISW signal peaks. These advances in observational data will help us make a better comparison between $\Lambda$CDM and evolving dark energy models in the coming years.

\begin{acknowledgements}
   The authors thank Kate Storey-Fisher, Giulio Fabbian, the \emph{Quaia} team for their help with the input quasar catalogue.
   
    The Large-Scale Structure (LSS) research group at Konkoly Observatory has been supported by a \emph{Lend\"ulet} excellence grant by the Hungarian Academy of Sciences (MTA). This project has received funding from the European Union’s Horizon Europe research and innovation programme under the Marie Skłodowska-Curie grant agreement number 101130774. Funding for this project was also available in part through the Hungarian National Research, Development and Innovation Office (NKFIH, grant OTKA NN147550). 
    M.G.Y. was supported by the EK\"OP-25 University Excellence Scholarship Program of the Ministry for Culture and Innovation from the source of the National Research, Development and Innovation Fund.
    IS acknowledges NASA grant 80NSSC24K1489.

    JRBC acknowledges support from a Momentum MSCA Fellowship, co-funded by the European Comission through the HORIZON-MSCA-2023-COFUND programme and the Secretariat of the Hungarian Academy of Sciences (MTA).
    
    We are grateful for the possibility to use HUN-REN Cloud (see \cite{Heder2022}; \url{https://science-cloud.hu/}), which helped us achieve the results published in this paper. Some of the results in this paper have been derived using the healpy and \healpix \, package.

\end{acknowledgements}

\bibliographystyle{aa}
\bibliography{refs}

\end{document}